\documentclass[final,1p,times]{elsarticle} 
\usepackage{graphicx} 
\usepackage{amssymb} 
\usepackage{amsthm} 
\usepackage{lineno} 

\journal{Nuclear Physics A} 
\begin{document} 

\begin{frontmatter} 

\title{Results from cosmics and first LHC beam with the
ALICE HMPID detector}

\author{G. Volpe$^{a}$ for the ALICE collaboration}

\address[a]{Dipartimento di Fisica dell'Universit$\grave{\textrm{a}}$ degli Studi di Bari and INFN, Sezione di Bari\\
         via E. Orabona 4, 70126 Bari, Italy.}

\begin{abstract} 
The ALICE HMPID (High Momentum Particle IDentification) detector has
been designed to identify charged pions and kaons in the range
1 $\div$ 3 GeV/c and protons in the range 1.5~$\div$~5~GeV/c.
It consists of seven identical proximity focusing RICH (Ring Imaging Cherenkov)
counters, covering in total 11 m$^2$, which exploit large area MWPC equipped with CsI
photocathodes for Cherenkov light imaging emitted in a liquid
C$_6$F$_{14}$ radiator. The ALICE detector has been widely
commissioned using cosmics and LHC beam from December 2007 until
October 2008. During the cosmics data taking the HMPID detector
collected a large set of data, using mainly the trigger provided by
the TOF detector. We
present here preliminary results of detector alignment using TPC
tracking. The HMPID could be operated in a stable way, at a safe HV
setting, also during LHC beam injection and circulation tests, when
a very large occupancy (up to 50\%) was achieved. Resulting gain
mapping and overall detector performance will also be discussed.

\end{abstract} 

\end{frontmatter} 


\section{ALICE-HMPID detector}
\begin{figure}
  \centering
  \includegraphics[scale=0.57]{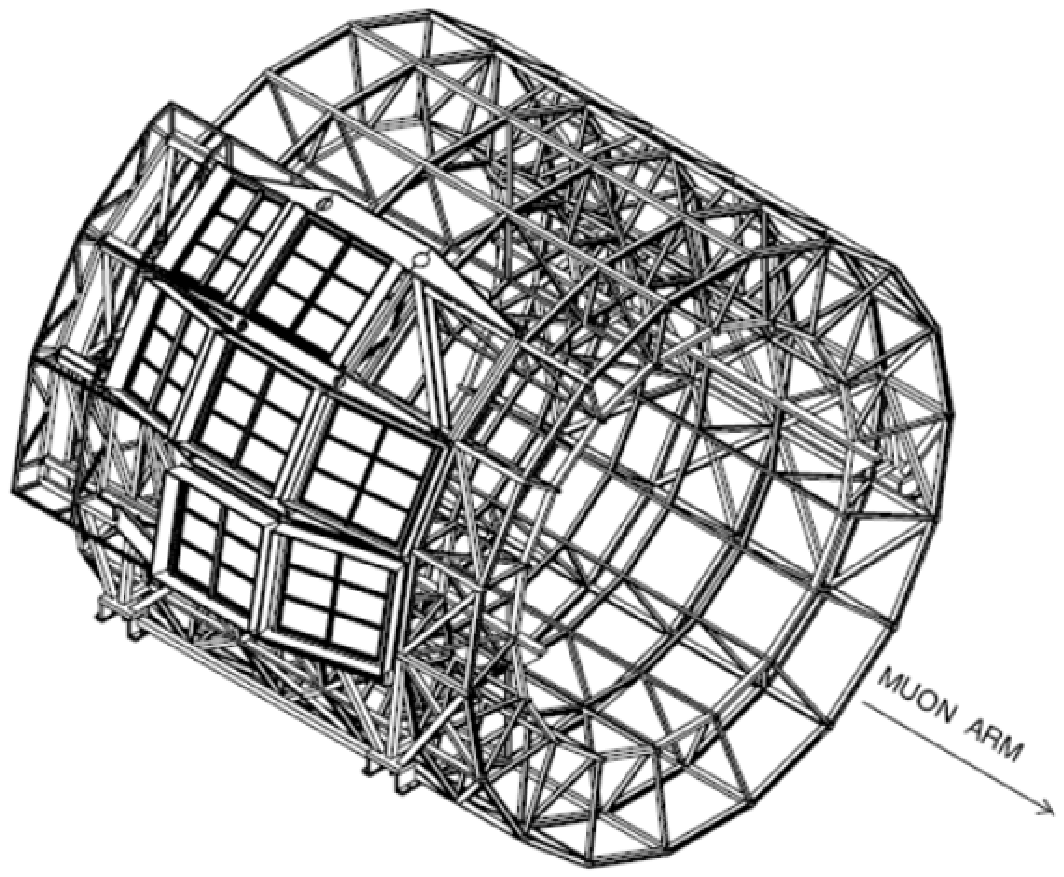}
  \includegraphics[trim= 25mm 70mm 20mm 0mm, clip, scale=0.42]{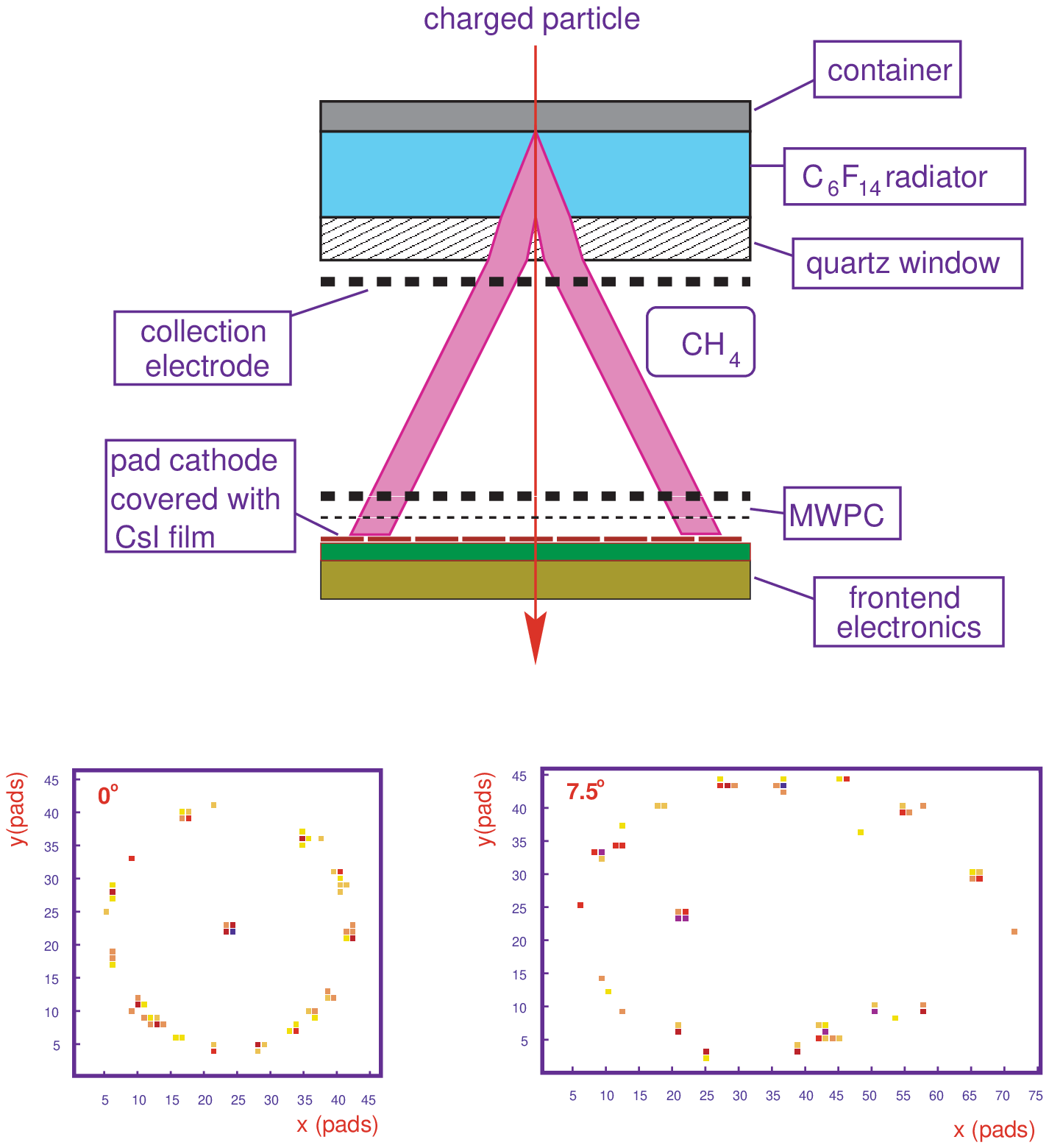}
  \vspace{-0.3cm}
  \caption{Left: schematic view of the ALICE-HMPID detector. Right: schematic section of one HMPID module.}
  \label{hmpid}
\end{figure}

The ALICE-HMPID \cite{tdr,cozza} (High Momentum Charged Particle
Identification Detector) performs charged particle track-by-track
identification by means of the measurement of the Cherenkov angle,
exploiting the momentum information provided by the tracking
devices. It consists of seven identical proximity focusing RICH
(Ring Imaging Cherenkov) counters. In Fig.~\ref{hmpid} is shown a
schematic view of one HMPID module. The radiator used
is C$_6$F$_{14}$ (n~$\approx$~1.2989 @ 175 nm, $\beta_{th}$~=~0.77),
15~mm thick. The photon detection is provided by multiwire chamber
coupled with pad-segmented CsI photocathode (CsI Q.E.~$\approx$~25\%
@ 175 nm, pads size 0.8x0.84 cm$^2$). The amplification gas is
CH$_4$ at atmospheric pressure, the anode-cathode gap is 2 mm and
the operational voltage is 2050~V (gain~$\approx$~4$\cdot$10$^4$).
The 42 photocathodes are segmented in 3840 pads with individual
analog readout. The Front-End and Readout electronics are
based on GASSIPLEX and DILOGIC chips developed within the HMPID
project. The noise level is 1~ADC channel (1000~e$^-$). HMPID performs charged hadron
identification in the intermediate
momentum region: 1$<p<$3 GeV/c for pions and kaons and 1.5$<p<$5 GeV/c for protons \cite{PPRI} and can contribute to:

\begin{itemize}
    \item Particle ratios vs. p$_T$ ($\bar{p}$/p, p/$\pi$, K/$\pi$);
    \item Jet physics: study jet fragmentation with the identification of particles in the jet and
    the "flavor" of the leading particle;
    \item Measurement of resonances production such as
    $\Phi_{(1020)}$ $\rightarrow$ K$^+$K$^-$;
    \item Light nuclei (alpha, deuteron, tritium) production.
\end{itemize}

\begin{figure}
  \centering
  \includegraphics[scale=0.4]{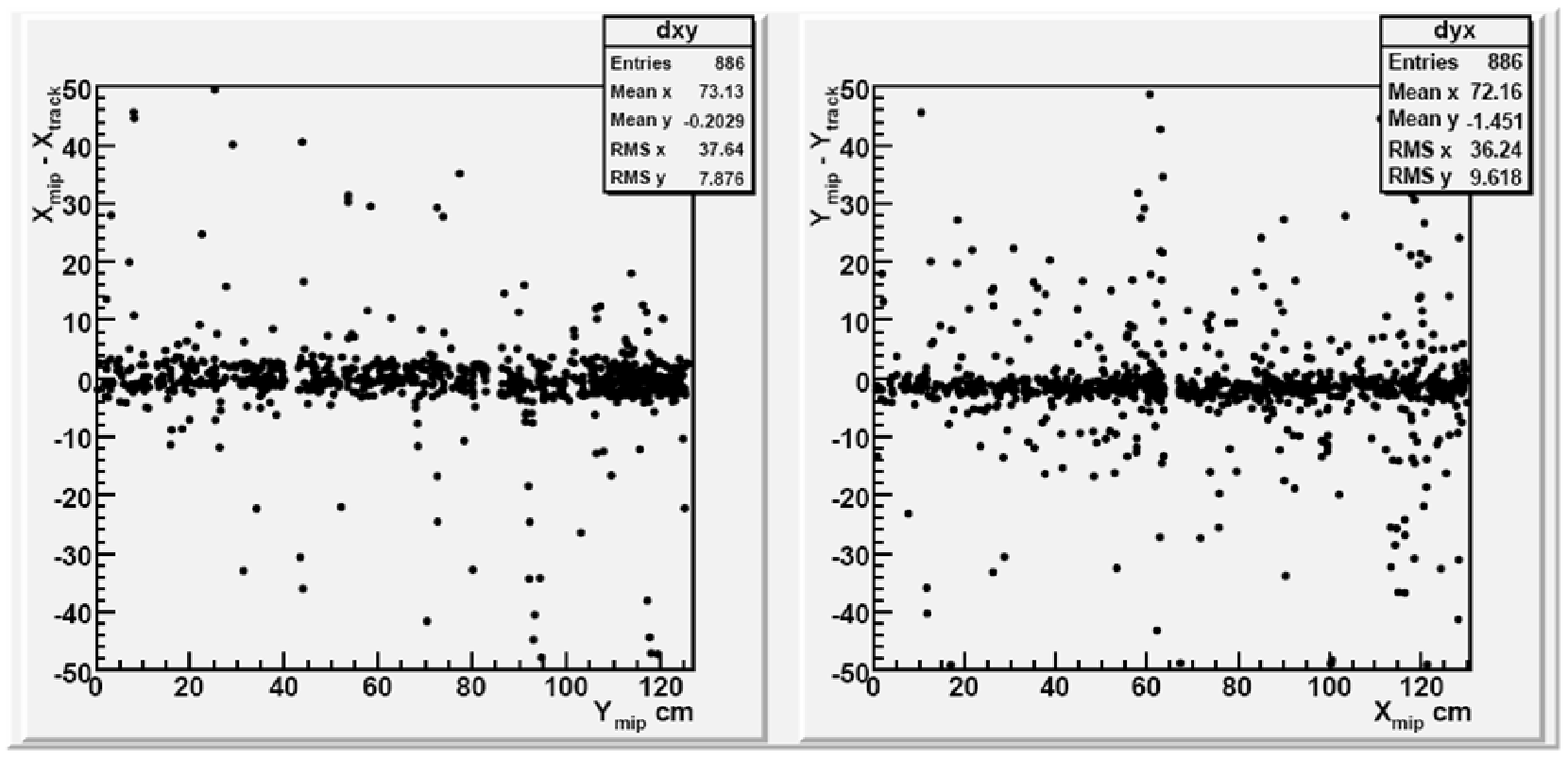}
  \includegraphics[scale=0.4]{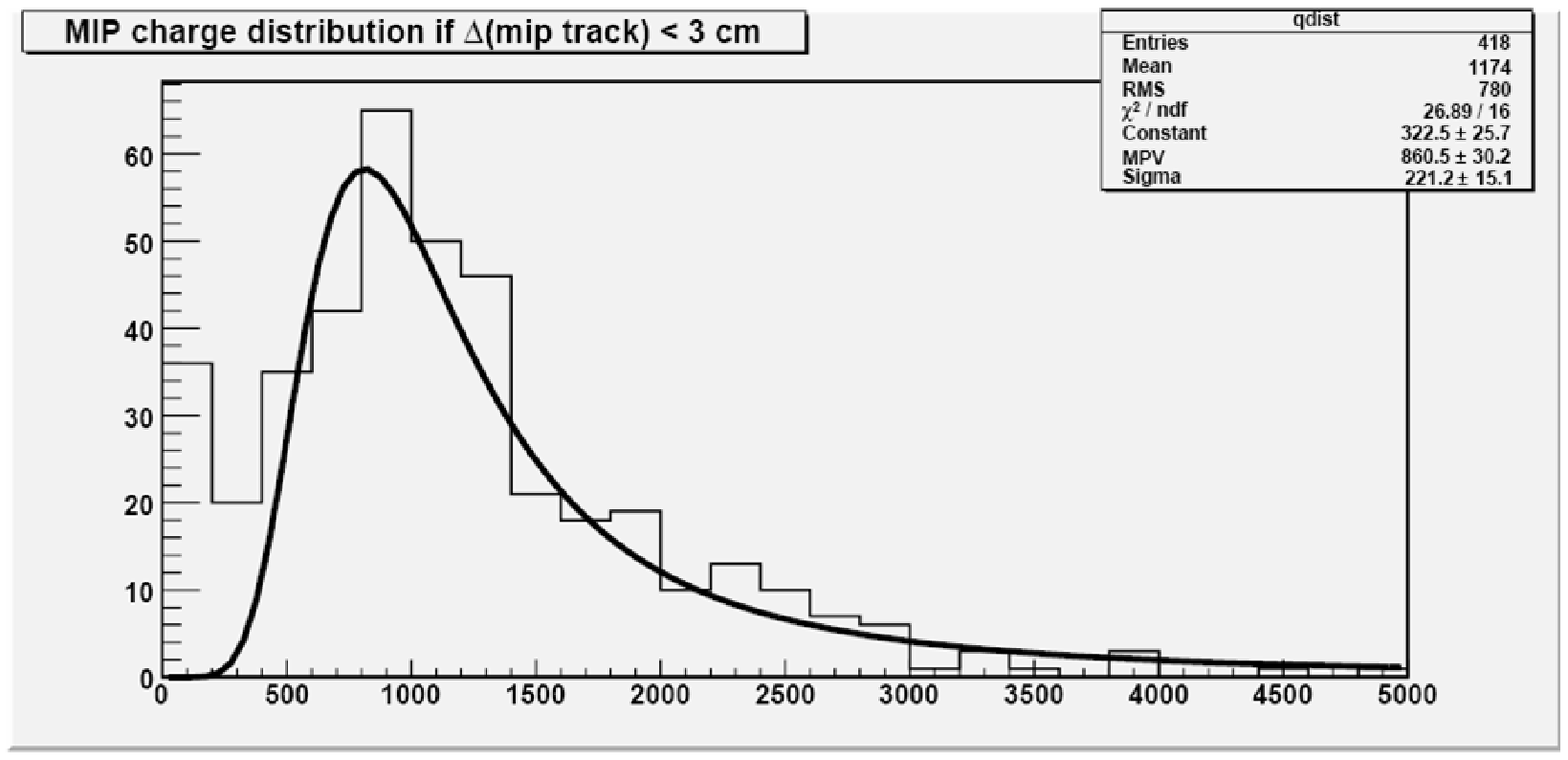}
  \caption{Left: track matching TPC-HMPID, no alignment procedures have been applied. Right: clusters
  charge distribution selecting clusters corresponding to a track in the TPC (Landau MPV $\approx$ 860 ADC).}
  \label{cosmics}
\end{figure}

\section{Results from cosmic}
\begin{figure}
  \centering
  \includegraphics[scale=0.18]{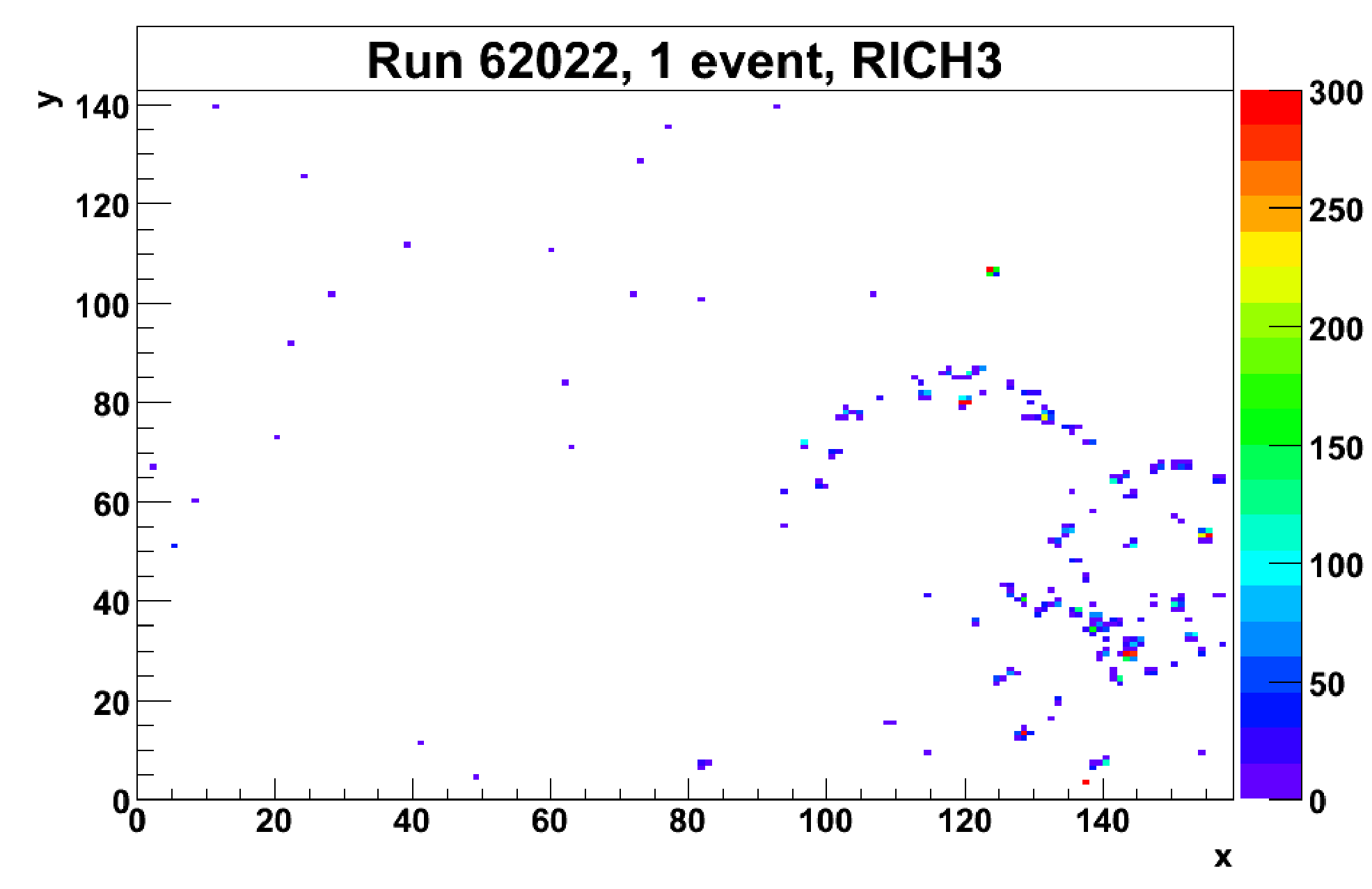}
  \includegraphics[scale=0.32]{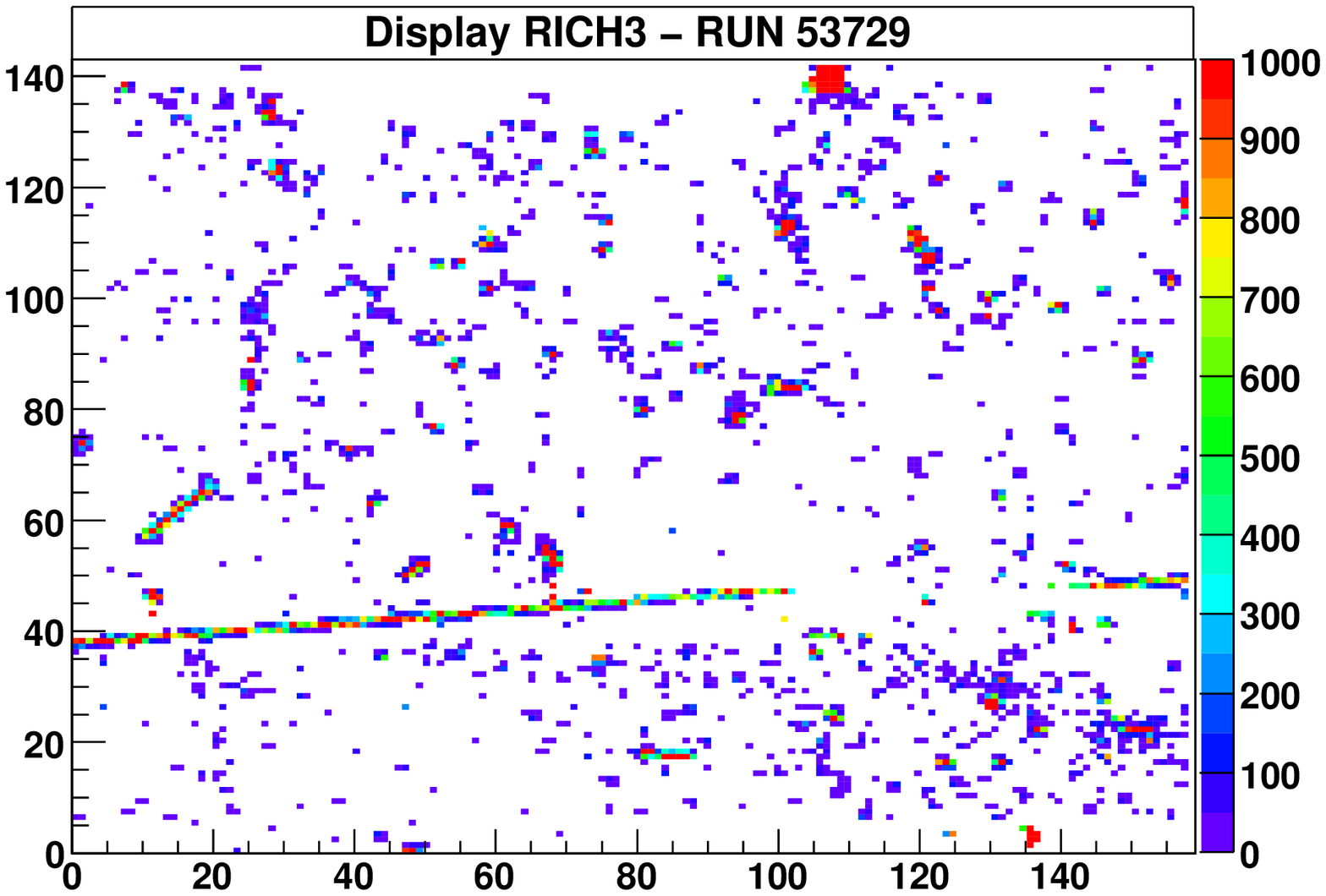}
  \caption{Left: display of HMPID module 3 for one cosmic event, cherenkov ring pattern is present.
  Right: display of HMPID module 3 for one event when the beam went through ALICE.}
  \label{display}
\end{figure}

A large set of data has been collected during the ALICE cosmic runs
allowing a first evaluation of detector performance and alignment, using the trigger provided by
the TOF (Time-of-Flight) detector, due to the most interesting trigger combination for such a kind of events.
The data analysis has been executed in AliRoot \cite{aliroot}, the
official off-line framework of the ALICE experiment.
In the left part of Fig.~\ref{cosmics} the difference between the x and y coordinate of
MIP cluster and the associated track impact point on photocathode vs
x$_{mip}$ and y$_{mip}$ respectively, are shown. The matching with
tracks reconstructed by means of Time-Projection-Chamber (TPC) is
evident, nevertheless alignment procedures have not been applied. In
the right part of Fig.~\ref{cosmics} the charge distribution of clusters corresponding
to a track in the TPC is shown. Despite the low statistics, the fit
with a Landau distribution is quite good. Because of the cosmic rays
spatial distribution, for the data taken with the trigger provided
by TOF the probability to have Cherenkov ring is
very low, anyway sometimes secondaries produced can generate
Cherenkov signal. Fig. \ref{display} shows the display for one event
of HMPID chamber 3, where a Cherenkov ring pattern is present.

\begin{figure}
  \centering
  \includegraphics[scale=0.32]{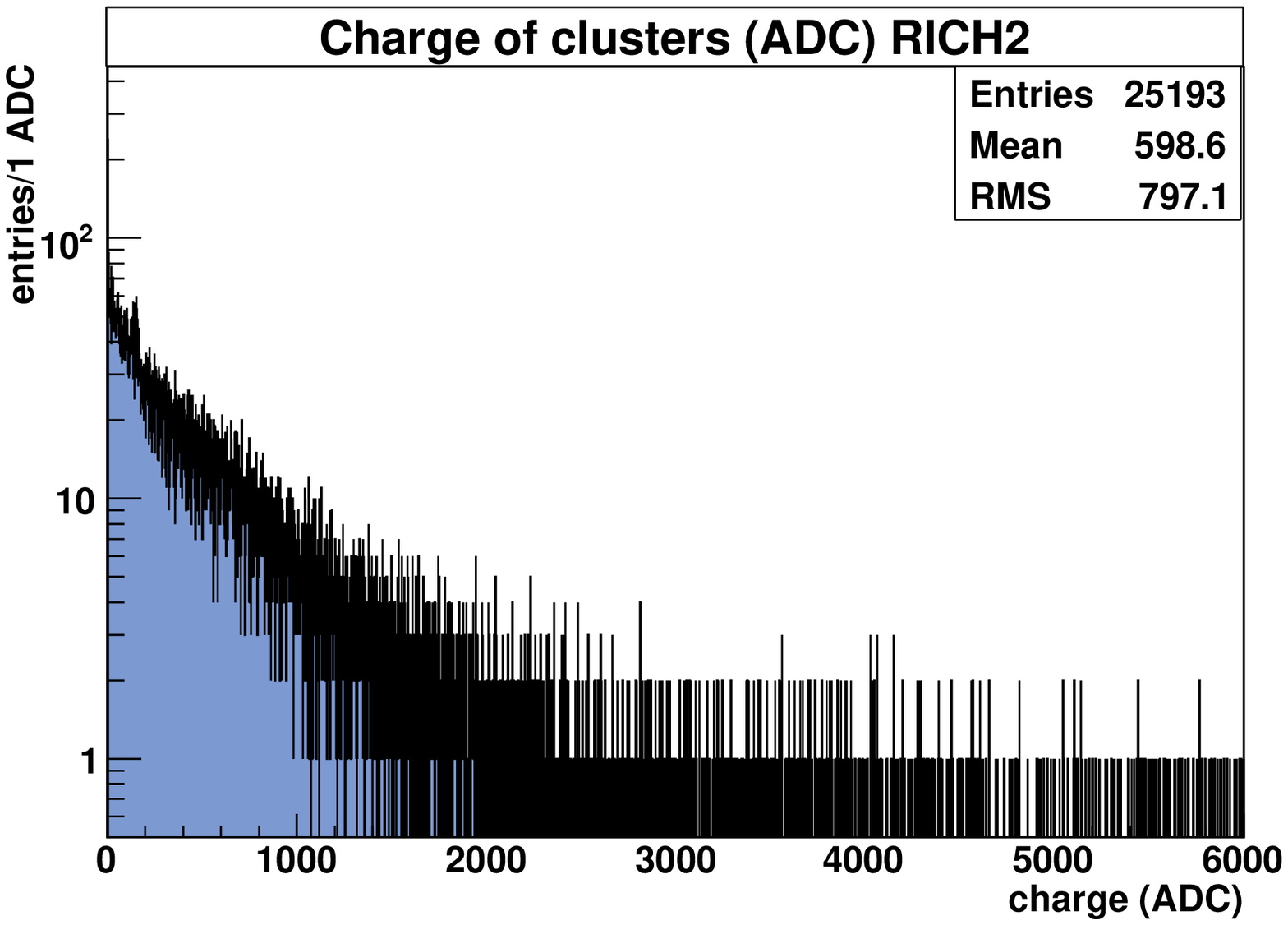}
  \includegraphics[scale=0.32]{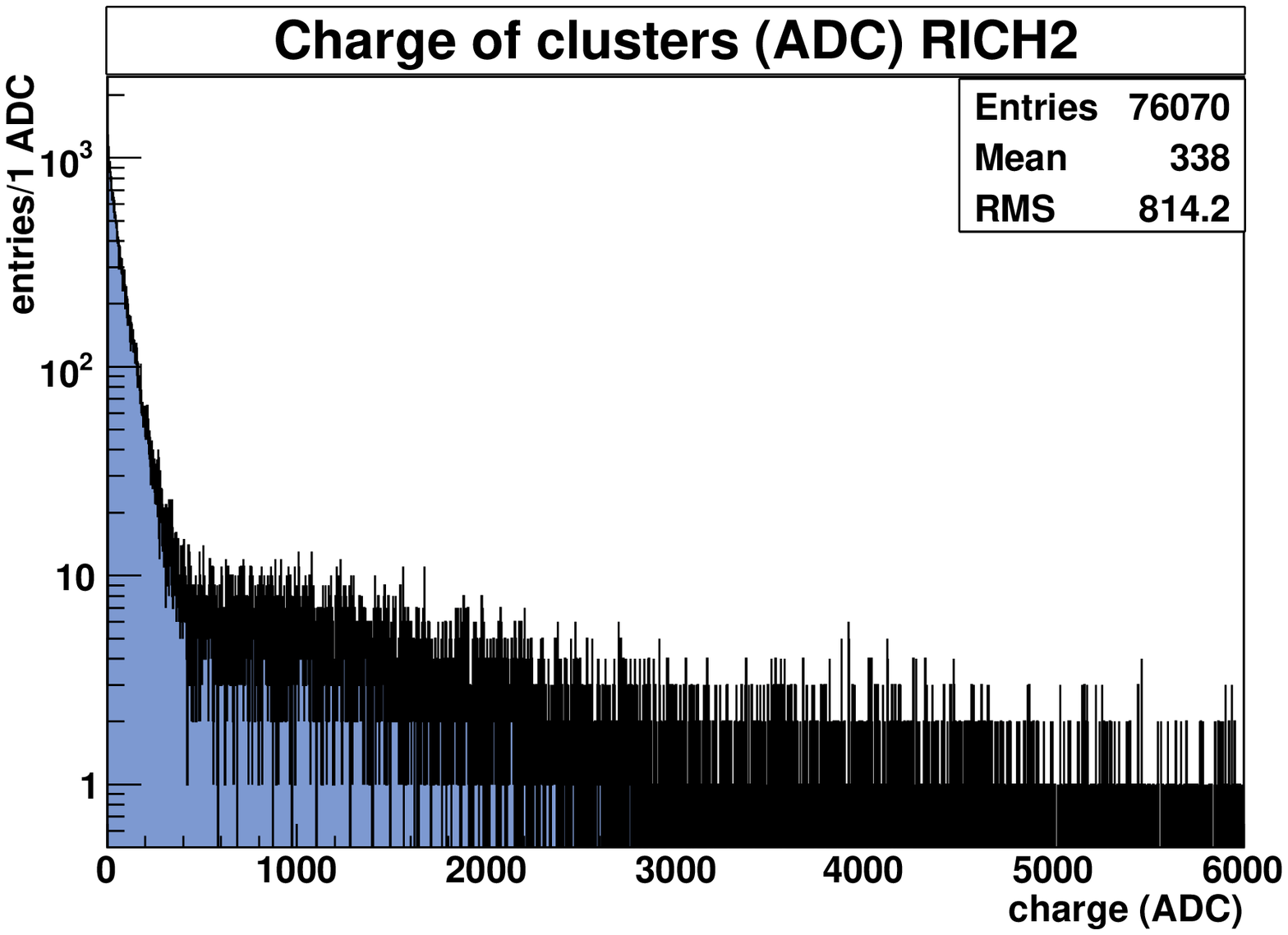}
  \caption{Left: total clusters charge distribution for chamber 2 for events from beam dump.
  Right: total clusters charge distribution for chamber 2 for events when the beam went through ALICE.}
  \label{charge}
\end{figure}

\begin{figure}
  \centering
  \includegraphics[scale=0.41]{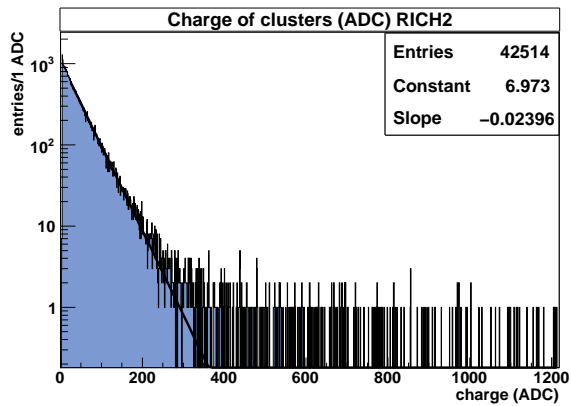}
  \vspace{-0.2cm}
  \caption{Clusters charge distribution selecting cluster with size less that 3. A$_0$ = -1/slope.}
  \label{charge2}
\end{figure}

\section{Results from LHC injection test and MWPCs gain estimation}
Data have been collected also during LHC beam injection tests.
Fig.~\ref{charge} shows the total cluster charge distribution for chamber
2. These data refer to high muon flux generated by the dump of the
LHC 450 GeV/c proton beam. Because of the high particle flux
expected the chamber voltage was set at 1860~V, to avoid trips. The
occupancy was $\approx$ 50\% on average. Fig.~\ref{display} shows
the display for one event of chamber 3. In this case the beam is not
dumped but it went through ALICE, and the chamber voltage was set at
the normal operational value. The occupancy for these events has
been of 12\% on average. In Fig.~\ref{charge} the total clusters charge
distribution for chamber 2 is shown. It is important to point out
that during the beam through ALICE, just beam-gas interaction events
were expected, therefore very low particle multiplicity. The high
particles multiplicity detected was due to beam screens present
along the beam path.

For a MWPC with the same characteristic of the HMPID one,
the single electron pulse height follows an exponential distribution \cite{tdr}.
From the data recorded when the beam went through ALICE, a rough estimate of the mean pulse height value (A$_0$)
has been executed. According to test beam studies, single
electron clusters have size less than 3 on avarage \cite{tdr}. In Fig. \ref{charge2}
the charge distribution of cluster with that selection for chamber 2 is shown.
In table 1, A$_0$ values in ADC units for all HMPID chambers are shown.
Indeed, the values of  the Landau distribution MPV (Fig.\ref{cosmics}) and A$_0$ retrieved,
do not correspond to the expected one at the HV values used. Investigation has been carried
out and an error in the calibration of HV board has been discovered.
The relation between the nominal HV value and
the real one is V$_{out}$ = V$_{set}$ + V$_{set}$*2.83*10$^{-2}$, that corresponds to a positive
bias of 58 V at 2050~V; according to this, the results obtained are consistent.

\begin{table}[h]
\begin{center}
\begin{tabular}{|c|c|} \hline
{module} & {A$_0$(ADC)}
\\ \hline {RICH0} & {46.9}
\\ \hline {RICH1} & {41.1}
\\ \hline {RICH2} & {41.2}
\\ \hline {RICH3} & {38.7}
\\ \hline {RICH4} & {43.5}
\\ \hline {RICH5} & {36.8}
\\ \hline {RICH6} & {47.1} \\ \hline
\end{tabular}
\end{center}
\centering\small\small{Table 1. \emph{A$_0$ values for the different
HMPID chambers}}
\end{table}

\section{Conclusion}
The results obtained from the cosmics and beam injection test have proven that HMPID works as expected.
Nevertheless calibration and alignment with cosmics are not
possible since high track statistics are necessary. A hardware problem has been detected and solved.
HMPID is waiting for collisions at LHC.

\end{document}